\documentstyle[12pt]{article}

\catcode`@=11
\def\secteqno{\@addtoreset{equation}{section}%
\def\theequation{\thesection.\arabic{equation}}}
\catcode`@=12
\secteqno
\newcommand{\be}{\begin{equation}}
\newcommand{\ee}{\end{equation}}
\newcommand{\bea}{\begin{eqnarray}}
\newcommand{\eea}{\end{eqnarray}}
\newcommand{\bref}[1]{(\ref{#1})}

\catcode`@=11

\begin{document}
\vfill
\vbox{
\today
%\hfill TOHO-FP-0066
\hfill 
}\null

\vskip 10mm
\begin{center}
{\Large\bf SO$(2,d-1)$ Gauge Theory of Gravity in $d$ Dimensional Spacetime and $AdS_d/CFT_{d-1}$ Correspondence.
\\
}\par
\vskip 10mm
{Takeshi FUKUYAMA\footnote{\tt E-mail:fukuyama@se.ritsumei.ac.jp}}\par
\medskip
Department of Physics and R-GIRO, Ritsumeikan University,\\ 
Kusatsu, Shiga, 525-8577 Japan\\
\medskip
\vskip 10mm
\end{center}
\vskip 10mm
\begin{abstract}
Gravity in d dimensions is formulated as the gauge theory of local SO(2,d-1) gauge group. The Chern-Pontryagin index ${\cal P}_{2n}$ plays a crucial role in both gravity and gauge theories. ${\cal P}_{2n}(gravity)$ gives the gravitational Lagrangian in 2n dimensions, having the vacuum solution $AdS_{2n}$. The same but global symmetry is shared with the gauge theories and 0,1-cochains of the Chern-Simon index ${\cal C}_{2n}(gauge)$ take part of $CFT_{2n-1}$ and $CFT_{2n-2}$, respectively. Gravity in odd dimensions is quite analogously formulated to that in even dimensions. This gives new insights on AdS/CFT correspondence.

\end{abstract}

%\vskip 2cm
%\newpage
%%%%%%%%%%%%%%%%%%%%%%%%%%%%%%%%%%%%%%%%%%%%%%%%%%%%%%%%%%%%%%%%%%%%%%%%

\section{Introduction}
Gauge theory of gravity has been old and long standing theme since the seminal work of Utiyama \cite{Utiyama1}.
In his work, spin connection $\omega_{\mu ab}$ was introduced as the gauge field of local Lorentz transformation, whereas the tetrad was treated as external field. The tetrad was incorporated into "gauge field" of the Poincare group
by Kibble \cite{Kibble}.
However, this is restricted rather in formal analogy and left unclear the question why the tetrad transform covariantly under the gauge transformation since the gauge fields $A_\mu$ transforms 
\be
A_\mu^U=U^{-1}(x)A_\mu U(x)+U^{-1}\partial_\mu U(x).
\ee
If we hope to incorporate the metric tensor or tetrad (vielbein, in general) in the usual gauge formulation, we must take the peculiar property of gravity into consideration, with which the usual gauge fields do not share certainly\footnote{The metric was introduced from invariance under general coordinate transformations \cite{F-U}}. One of the most important points is the soldering of internal space of gauge symmetry of gravity with the external space. This has no analogy in the other Abelian and non-Abelian gauge fields.
So the gauge theory of gravity must reflects this peculiarity.
As we will show in this letter, SO(2,3) gauge group satisfies these properties.
We proposed the gauge theory of gravity in the symmetry breking chain of
the conformal (SO(2,4)) $\rightarrow$ SO(2,3) \cite{Fukuyama}.

The conformal transformation,
\be
dx_a'dx_a'=\left(\Omega(x)\right)^{-2}dx_adx_a ~~~(a=1,..,d),
\label{cft}
\ee
in flat (1,d-1) dimensions corresponds to the "enlarged" SO(2,d) symmetry which maps
\be
\sum_1^{d+2} Z_A^2=0
\ee
into itself, where two $Z_A$ components are timelike.
That is,
\be
Z_A'=\Lambda_A{}^BZ_B~~ \mbox{with}~~\Lambda_A{}^C\Lambda_C{}^B=k(Z)\delta_A^B.
\ee
There appears no scale parameter and SO(2,d) is the maximal spacetime symmetry of massless fields in d dimensional spacetime. We asserted that gravity is formulated as the gauge theory of local SO(2,d-1), leaving 
\be
\sum_1^{d+1} Z_A^2=-l^2.
\ee
invariant \cite{Fukuyama} \cite{MacDowell} \cite{Stelle}. Real $l$ measures the scale breaking from SO(2,d) with two timelike $Z_A$.
Its vacuum solution $AdS_d$ has SO(2,d-1) symmetry, which is also global symmetry of $CFT_{d-1}$.

This fact reminds us the AdS/CFT correspondence \cite{Maldacena}\cite{Oguri}.

However, it is a little bit curious to us who has studied the gauge theory of gravity since old and not so popular era. For gravity is also one of the gauge theories as mentioned above and there arises an expectation that the gauge theoretical view point of gravity may shed new light on AdS/CFT and vice versa\footnote{We do not discuss on the supersymmetry in this letter.}.

This paper is organized as follows.
In section 2, we review gravity of SO(2,3) gauge group \cite{Fukuyama} corresponding to $d=4$.
Arguments are extended to general $d$ in section 3. $d=2$ has peculiar property and separately discussed in section 4.
In these constructions, the topological objects of the Chern-Pontryagin class and Chern-Simon class concerned with the symmetry SO(2,d) play crucial roles in both gravity and gauge theories.
Section 5 is devoted to the discussions.

\section{SO(2,3) Gauge Theory of Gravity}
Before discussing SO(2,3) invariance we will very briefly review the relation between the conformal transformation \bref{cft} and SO(2,d) \cite{Dirac}.
We will describe conformally flat (1,d-1) coordinates as $x^a$.
The infinitesimal conformal transfomation 
\be
x'^a=x^a+v^a(x)
\ee
must satisfy
\be
\partial_av_b+\partial_bv_a=\frac{2}{d}\delta_{ab}.
\label{CF1}
\ee
The general solution to this equation is
\be
v_a=p_a+\omega_{ab}x^a+\lambda x_a+f_ax^2-2x_af_bx^b.
\label{cft2}
\ee
This symmetry is described by the SO(2,d) invariant transformation acting on
%of \bref{cft} is
%We shortly review SO(2,3) gauge theory of gravity \cite{Fukuyama}\cite{MacDowel%l}\cite{Stelle}, where tetrad is defined by $h_{\mu a}=\omega_{\mu 5a}$  where %$\omega_{\mu 5a},~\omega_{\mu ab}~ (a,b=1,...,4)$ are the connections of SO(2,3%) gauge group,
%\be
%Z_A(x)^2=Z_1^2+...+Z_4^2+Z_5^2=-l^2.
%\ee
%Here $l$ is the scale breaking parameter from the $SO(2,4)$ acting on
six components projective coordinates $Z_A$ 
\be
Z_A=(Z_a, Z_-, Z_+)~~(a=1,..,d)
\ee
with the constraints
\be
Z_A^2=Z_aZ_b\eta^{ab}-2Z^+Z^-=0,
\label{cft3}
\ee
where $Z_\pm=Z_{d+1}\pm Z_{d+2}$ for timelike $Z_{d+1}$.
Indeed if we define 
\be
x^a=\frac{Z^a}{Z^+},
\ee
 SO(2,d) generators $G^{AB}=Z^{[A}p^{B]}$ produce \bref{cft2}.
Gauge group of gravitation represents the symmetry of the spacetime and naively we might consider the conformal symmetry as the gauge group.
We spontaneously break the conformal invariance by
\be
(Z_{d+1},Z_{d+2})=(l,l)~~\mbox{or}~~(-l,-l).
\label{cft4}
\ee
However, it needs a dimensional parameter $l$, which breaks conformal symmetry explicitly. 
Thus we considered the gauge theory of gravity whose gauge group is reduced from SO(2,d) to SO(2,d-1) \cite{Fukuyama} \footnote{The implication of conformal invariance in gravity is further discussed in section 5.}. In this section, hereafter we consider $d=4$ and 
\be
Z_1^2+...+Z_4^2+Z_5^2=-l^2.
\ee
From gauge theoretical view point we may put SO(1,4) in place of SO(2,3) with $l^2$ in place of $-l^2$. However, we consider SO(2,3) taking supergravity into consideration. (The extra dimension must be compactified and in total ten dimensions, 
cosmological constant must be zero.)
Correspondng to SO(2,3), the covariant derivative is defined by
\be
D_\mu=\partial_\mu-i\omega_{\mu AB}S_{AB}/2~~(A,B=1,...,4,5)
\ee
Here $\omega_{\mu AB}$ are $4\times 10$ connection fields and $S_{AB}$ are the generators of (anti) de Sitter group.
The field strength is derived from the commutation relation
\be
i[D_\mu,D_\nu]=-R_{\mu\nu AB}S_{AB}/2.
\ee
\be
R_{\mu\nu AB}=\partial_\mu\omega_{\nu AB}-\partial_\nu\omega_{\mu AB}-\omega_{\mu AC}\omega_{\nu CB}+\omega_{\nu AC}\omega_{\mu CB}
\ee
The Einstein's action is written as
\be
I=\int d^4x \epsilon^{ABCDE}\epsilon^{\mu\nu\rho\sigma}(Z_A/l)\left[R_{\mu\nu BC}R_{\rho\sigma DE}/(16g^2)+ D_\mu Z_BD_\nu Z_CD_\rho Z_DD_\sigma Z_E\sigma (x)\{(Z_A^2/l^2)-1\}^2 \right].
\label{action1}
\ee

Here $\epsilon^{\mu\nu\lambda\sigma}$ and $\epsilon^{ABCDE}$ are fully antisymmetric tensors with $\epsilon^{1234}=1$ and $\epsilon^{12345}=1$, respectively ($4,5$ are timelike components).
It should be remarked that this action is a geometrical invariant and that we do not introduce metric ad hoc. Hamilton formulation of action \bref{action1} was given by \cite{fuku-kami}.
After the gauge choice
\be
Z^A=(0,0,0,0,l)
\label{GF1}
\ee
\be
D_\mu Z_A=(\partial_\mu\delta_{AB}-\omega_{\mu AB})Z_B =\{\begin{array}{cc}\omega_{\mu a5}l\equiv e_{\mu a} & \mbox{if}~ A=a\\
  0 & \mbox{if}~ A=5 \end{array} .
\label{tetrad1}
\ee
It is important that $e_{\mu a}$ transforms covariantly under the remaining 4-dim Lorentz rotation.
Generalized Riemannian tensor $R_{\mu\nu ab}$ is divided into two terms
\be
R_{\mu\nu ab}=\mathring{R}_{\mu\nu ab}-e_{[\mu a}e_{\nu]b}/l^2.
\ee
Here $\mathring{R}_{\mu\nu ab}$ is the conventional Riemannian tensor defined by
\be
\mathring{R}_{\mu\nu ab}=\partial_{[\mu}\omega_{\nu]ab}-\omega_{[\mu ac}\omega_{\nu]cb}
\ee
and $e_{[\mu a}e_{\nu]b}\equiv e_{\mu a}e_{\nu b}-e_{\nu a}e_{\mu b}$.

$L_{grav}$ takes the form
\bea
L_{grav}&=&{\cal P}_4(gravity)= \epsilon^{abcd}\epsilon^{\mu\nu\rho\sigma}R_{\mu\nu ab}R_{\rho\sigma cd}/(16g^2)\nonumber\\
&=& \partial_\mu{\cal C}_4^\mu-e\left(\mathring{R}-\frac{6}{l^2}\right)/(16\pi G),
\label{4D2}
\eea
where 
\be 
16\pi G\equiv g^2l^2
\ee
\be
e=\mbox{det}e_{\mu a},~~\mathring{R}_{\mu a}=e^{\nu b}\mathring{R}_{\mu\nu ab},~~\mathring{R}=e^{\mu a}\mathring{R}_{\mu a},
\ee
 and use has been made of
\bea
\epsilon^{abcd}\epsilon^{\mu\nu\rho\sigma}e_{\mu a}e_{\nu b}e_{\rho c}e_{\sigma d}&=&4! e\nonumber\\
\epsilon^{abcd}\epsilon^{\mu\nu\rho\sigma}e_{\mu a}e_{\nu b}&=& 2e~e^{[\rho c}e^{\sigma] d}~~\mbox{etc.}
\eea
Here $e^{\mu a}e_{\mu b}=\delta_{ab},~e^{\mu a}e_{\nu a}=\delta^\mu_\nu$.
The quadratic term in $\mathring{R}_{\mu\nu ab}$ is total derivative $\partial_\mu {\cal C}_4^\mu$ (the Gauss-Bonnet term). 
This gauge theoretical construction of gravity may shed new light on the 
AdS/CFT correspondence \cite{Maldacena} \cite{Oguri} which states the correspondence between $AdS_d$ gravity and (d-1)dimensional cnformal field theory. 
Indeed, the Gauss-Bonnet term in \bref{4D2} does not affect the equation motion but does the boundary like event horizon of Black Hole (BH). So this may change the scenario of near-horizon extreme BH like Reissner-Nordstrom and Kerr etc. \cite{Strominger}. This is indeed the case and the special combinations of \bref{4D2} gives the conserved mass and angular momentum for Kerr-AdS BH \cite{Zanelli}.
As we will show in the next section, we have higher derivative terms other than the linear Einstein and cosmological terms. These also modify BH solution.

Three dimensional action of Yang-Mills gauge field $F_{\mu\nu}^aT^a\equiv F_{\mu\nu}$,
\be
S=\int d^3x \frac{-1}{2g^2}\mbox{Tr}F_{\mu\nu}F^{\mu\nu}.
\ee
has dimensional coupling (mass dimension of $g$ is $1/2$) and not SO(2,3) invariant. In the above arguments, $AdS_4$ has been derived with 4-dimensional Chern-Pontryagin index ${\cal P}_4(gravity)$. Unlike the SO(2,3) invariant gravity, the corresponding counterpart in gauge theory ${\cal P}_4(gauge)$ is total derivative, being related with 3 dimensional Chern-Simon term as follows.
\bea
L_{gauge}&=&{\cal P}_4(\mbox{gauge})=-\frac{1}{32\pi^2}\mbox{Tr}\epsilon^{\mu\nu\rho\sigma}F_{\mu\nu}F_{\rho\sigma}\label{4Dgauge}\nonumber\\
&=& \partial_\mu {\cal C}_4^\mu(\mbox{gauge}),
\eea
where
\be
{\cal C}_4^\mu(\mbox{gauge})=-\frac{1}{8\pi^2}\mbox{Tr}\epsilon^{\mu\nu\rho\sigma}\left(A_\nu\partial_\rho A_\sigma+\frac{2}{3}A_\nu A_\rho A_\sigma\right).
\label{CS1}
\ee

${\cal C}_4^\mu$ lives in one dimension lower than the Chern-Pontryagin ${\cal P}_4$ \cite{Jackiw2}\cite{Eguchi}, and we may consider
\be
{\cal C}_4^{(0)}(A)=-\frac{1}{8\pi^2}\mbox{Tr}\epsilon^{\nu\rho\sigma}\left(A_\nu\partial_\rho A_\sigma+\frac{2}{3}A_\nu A_\rho A_\sigma\right).
\label{4D}
\ee
${\cal C}^{(0)}(A)$ may be called a 0-cochain \cite{Jackiw2}\cite{Eguchi} and
\bea
{\cal C}_4^{(0)}(A^U)&=&{\cal C}_4^{(0)}(A)+\frac{1}{8\pi^2}\mbox{Tr}\epsilon^{\nu\rho\sigma}\partial\left(a_\rho A_\sigma\right)+\frac{1}{24\pi^2}\mbox{Tr}\epsilon^{\nu\rho\sigma}a_\mu a_\nu a_\sigma\nonumber\\
&\equiv& {\cal C}_4^{(0)}(A)+\Delta{\cal C}_4^{(0)},
\eea
where $a_\nu=\partial_\nu U U^{-1}$. $\Delta{\cal C}^{(0)}$ is also total derivative
,
\be
\Delta {\cal C}_4^{(0)}=\partial_\mu {\cal C}_4^{(1)\mu}(A,U)
\ee
and we obtain one-cochain
\be
{\cal C}_4^{(1)}(A,U)=\frac{1}{8\pi^2}\mbox{Tr}\epsilon^{\alpha\beta}a_\alpha A_\beta+\frac{d^{-1}}{24\pi^2}\mbox{Tr}\epsilon^{\alpha\beta\gamma}a_\alpha a_\beta a_\gamma.
\ee
Here $d^{-1}$ is simbolic of integral and the explicit form is given for the specfic case of SU(2) gauge in \cite{Jackiw2}. Thus \bref{4D} is gauge invariant up to topological winding number.
The dimensional descent is continued further:
\bea
\Delta {\cal C}_4^{(1)}&=&{\cal C}_4^{(1)}(A^{U_1};U_2)-{\cal C}_4^{(1)}(A;U_{12})+{\cal C}_4^{(1)}(A;U_1)\label{1-cochain}\\
&=&\partial_\mu{\cal C}_4^{(2)\mu}(A;U_1,U_2)
\eea
The explicit form of 2-cochain ${\cal C}^{(2)\mu}(A;U_1,U_2)$ is given in \cite{Cronstrom}

Thus ${\cal P}_4 (grav)$ and ${\cal C}_4^{(0)} (gauge)$ are correspondents of $AdS_4/CFT_3$. 
They are related with SO(2,3): In the former, it is local gauge group of gravity, and in the latter it is global symmetry of gauge theory.
${\cal C}_4^{(1)}(A,U)$ is related with $CFT_2$ as will be shown in the next section.

In the following sections, this is generalized to ${\cal P}_{2n} (grav)$ and ${\cal C}_{2n} (gauge)$
for $d=2n$. $d=2n+1$ cases are also discussed. Though gravitational part for $d=2n+1$ is not described as such topological object unlike that of $d=2n$, procedures are quite analogous to $d=2n$ cases.

\section{SO(2,d-1) Gravity for $d\neq 2$}
We have started with $d=4$ dimensional spacetime with SO(2.3) gauge group.
This formulation is easily extended to $d=5$, five dimensional spacetime.
That is
\bea
I&=&-\int d^5x \epsilon^{ABCDEF}\epsilon^{\mu\nu\rho\sigma\lambda}(Z_A/l)D_\mu Z_B\left[R_{\nu\rho CD}R_{\sigma\lambda EF}/(48g^2l)\right.\nonumber\\
&+&\left.D_\nu Z_CD_\rho Z_DD_\sigma Z_ED_\lambda Z_F \sigma (x)\sum_{A=1}^6\{(Z_A^2/l^2)-1\}^2 \right].
\label{5D}
\eea
with 
\be
Z_A=(0,0,0,0,0,l).
\label{GF2}
\ee
In this case
\be
D_\mu Z_A=(\partial_\mu\delta_{AB}-\omega_{\mu AB})Z_B =\{\begin{array}{cc}\omega_{\mu a6}l=e_{\mu a} & \mbox{if}~ A=a\\
  0 & \mbox{if}~ A=6 \end{array} ,
\ee
Here $\mu$ and $a$ run over 1,..,5 in world and local Lorentz coordinates, respectively.
Consequenly \bref{5D} is reduced to
\bea
L_{grav}&=& \epsilon^{abcde}\epsilon^{\mu\nu\rho\sigma\lambda}e_{\mu a}R_{\nu\rho bc}R_{\sigma\lambda de}/(48g^2l)\nonumber\\
&=&\epsilon^{abcde}\epsilon^{\mu\nu\rho\sigma\lambda}e_{\mu a}\mathring{R}_{\nu\rho bc}\mathring{R}_{\sigma\lambda de}/(48g^2l) -e\left(\mathring{R}-\frac{10}{l^2}\right)/(16\pi G_5)
\label{d3}
\eea
with $16\pi G_5=g^2l^3$.
Thus we obtain $AdS_5$ as the vacuum solution. In this case, however, higer derivative terms (the first term of \bref{d3}) are not total derivatives and change the equation of motion in high energy region and do therefore Black Hole solution and its near horizon property.

Corresponding to SO(2,4) in gauge theory is
\be
S_{gauge}=-\frac{1}{2g^2}\int d^4x \mbox{Tr}F_{\mu\nu}F^{\mu\nu}+\int d^4x {\cal P}_4(gauge),
\label{d4}
\ee
where ${\cal P}_4(gauge)$ is defined in \bref{4Dgauge}. \bref{d3} and \bref{d4} constitute the correspondents in $AdS_5/CFT_4$ at least from the invariance property.

In six dimensional spacetime, gauge group of gravity is SO(2,5), and gravity action is
\be
L_{gravity}=-\epsilon^{ABCDEFG}\epsilon^{\alpha\beta\mu\nu\rho\sigma}\left[(Z_A/l)R_{\alpha\beta BC}R_{\mu\nu DE}R_{\rho\sigma FG}\right].
\label{6D}
\ee
Further processes follow anologously to the preceeding arguments.
By the gauge choice $Z_A=(0,0,0,0,0,0,l)$, \bref{6D} becomes
\be
L_{grav}= \partial_\mu {\cal C}_6^\mu(\mathring{R})+\mbox{quadratic of} ~\mathring{R}_{..}-e\mathring{R}+\mbox{cosmological const}.
\ee
Thus we obtain the Einstein equation with negative cosmological constant in low energy. However, it includes terms quadratic in Riemannian tensor and Pontrjagin ${\cal C}_6(\mathring{R})$ term.
The corresponding counterpart of gauge action is
\bea
L_{gauge}&=&\frac{1}{384\pi^3}\epsilon^{\alpha\beta\mu\nu\rho\sigma}\mbox{Tr}F_{\alpha\beta}F_{\mu\nu}F_{\rho\sigma}\\
&=&\partial_\sigma {\cal C}_6^\sigma (gauge)
\eea
with
\be
{\cal C}_6^\sigma\equiv \frac{1}{192\pi^3}\epsilon^{\sigma\alpha\beta\mu\nu\rho}\mbox{Tr}\left(F_{\alpha\beta}F_{\mu\nu}A_\rho-F_{\alpha\beta}A_\mu A_\nu A_\rho+\frac{2}{5}A_\alpha A_\beta A_\mu A_\nu A_\rho\right).
\ee
and 
\be
{\cal C}_6^{(0)}=\frac{1}{192\pi^3}\epsilon^{\alpha\beta\mu\nu\rho}\mbox{Tr}\left(F_{\alpha\beta}F_{\mu\nu}A_\rho-F_{\alpha\beta}A_\mu A_\nu A_\rho+\frac{2}{5}A_\alpha A_\beta A_\mu A_\nu A_\rho\right).
\ee
Thus ${\cal P}_6(gravity)$ and ${\cal C}_6^{(0)}(gauge)$ are related with SO(2,5). We may add kinetic terms using scalar $\phi$ of mass dimension 1 and $CFT_5$ is,
\be
\phi (\partial \phi)^2+\phi F_{\mu\nu}^2+{\cal C}_6^{(0)}(gauge).
\ee
${\cal C}_6^{(1)}(gauge)$ defined analogously to \bref{1-cochain} may be added into \bref{d4}.

For $d=7,~ AdS_7$ is straightforward and omit to describe it.
The counterpart of gauge theories are guided by renormalizability
\cite{Seiberg}
\be
\Phi F_{\mu\nu}^2+B\wedge F\wedge F+(\partial\Phi)^2+(dB)^2+...,
\label{7D}
\ee
where a scalar $\Phi$ and a two form $B_{\mu\nu}$ both of dimension 2.
Unlike for $d\leq 5$, we have no idea to derive \bref{7D} from the dimensional descent of higher ${\cal C}_d$ since gauge field $B_{\mu\nu}$ has 2 dimensional world volume peculiar to superstring or supergravity.

The same procedures are performed for $d=3$ dimensinal case, SO(2,2) gravity.
\be
I=\int d^3x \epsilon^{ABCD}\epsilon^{\mu\nu\rho}(Z_A/l)D_\mu Z_B\left[R_{\nu\rho CD}/(2g^2l)+ D_\nu Z_CD_\rho Z_D\sigma (x)\sum_{A=1}^4\{(Z_A^2/l^2)-1\}^2 \right].
\label{3D}
\ee
After the gauge fix $Z_A=(0,0,0,l)$, \bref{3D} is reduced to 

\bea
S_{grav}&=& -\int d^3x \epsilon^{abc}\epsilon^{\mu\nu\rho}e_{\mu a}R_{\nu\rho bc}/(2g^2l)\nonumber\\
&=&-\int d^3x e\left(\mathring{R}-\frac{6}{l^2}\right)/(16\pi G_3),
\eea
where $g^2l=16\pi G_3$. The SO(2,2) invariant gauge theory is naively
\be
I_{gauge}=-\frac{1}{2g^2}\int d^2x \epsilon^{\mu\nu}F_{\mu\nu}=\int d^2x {\cal P}_2(gauge)
\ee
but it is topological invariant.
The corresponding gauge counterpart is the WZW model \cite{WZW}
\be
I_{WZW}=\frac{1}{4\pi}\int d^2x Tr(a_za_{\overline{z}})+{\cal C}_4^{(1)}(A,U).
\ee
Here $a_\nu$ is defined at (2.19) with $z=x-it$.
${\cal C}_4^{(1)}(A,U)$ is given by (2.27).

We have extended our formulation to $d=3,~5,~6,~7$.
$d=2$ case is discussed in the next section.

%This is straightforwardly extended to arbitrary even and odd dimensions for $d>%2)$.
%As you can easily understood, the situations are difficult in even and odd dime%nsions conceptually. However, the technical procedures are quite analogous in b%oth cases.

\section{SO(2,1) Gravity}
Two dimensional gauge theory of gravity is special in the sence that its gauge group SO(2,1) has an infinite set of generators,
\bea
L_m&=&\frac{T}{2}\int_0^\pi e^{-2im\sigma}T_{--}d\sigma~~~(m:\mbox{integer})\nonumber\\
\bar{L}_m&=&\frac{T}{2}\int_0^\pi e^{2im\sigma}T_{++}d\sigma.
\eea
These generators satisfy the Virasoro algebra
\be
[L_m, L_n]=(m-n)L_{m+n}+\frac{c}{12}(m^3-m)\delta_{m,-n}.
\ee
Usually this corresponds to the fact that 
\be
S=-\int d^2x\sqrt{-g}(R-\Lambda)
\label{2dG}
\ee
is total derivative. The following non trivial action 
was proposed by Jackiw and Teitelboim \cite{Jackiw}. 
\be
S=-\int d^2x\sqrt{-g}(R-\Lambda)N
\label{Jackiw}
\ee
with auxiliary field $N$. This action was formulated as the SO(2,1) gauge theory of gravity by us \cite{Fukuyama2}.
Naively we might consider
\be
S=-\int d^2x \epsilon^{ABC}\epsilon^{\mu\nu}R_{\mu\nu AB}Z_C/l.
\ee
Unfortunately, this leads us to \bref{2dG}.  

However, two dimensionality has the peculiar property that the scalar fields (we denote them as $\phi_A$) have the canonical dimensionality 0, which makes us possible to construct
\be
S=-\frac{1}{2}\int d^2x \epsilon^{ABC}\epsilon^{\mu\nu}R_{\mu\nu AB}\phi_C.
\label{2dG2}
\ee
The equations of motion derived from \bref{2dG2} are
\bea
R_{AB}&=&d\omega_{AB}-\omega^2_{AB}=0,\label{2deq1}\\
D\phi_A&=& d\phi_A-\omega_{AB}\phi_B=0\label{2deq2},
\eea
where $R_{AB}=\frac{1}{2}R_{\mu\nu AB}dx^\mu\wedge dx^\nu,~\omega_{AB}=\omega_{\mu AB}dx^\mu$.
By decomposing \bref{2deq1} into (0,1) and (a,2) components, we obtain
\bea
0&=& de_a-\omega_{ab}e_b ~~~~~~(a,b=0,1),\\
0&=& d\omega_{01}-e_0e_1/l^2.
\label{2deq5}
\eea
In the same way, \bref{2deq2} gives
\bea
0&=&d\phi_a-\omega_{ab}\phi_b+e_a\phi_2/l,\label{2deq3}\\
0&=&d\phi_2+e_a\phi_a/l.\label{2deq4}
\eea
\bref{2deq4} is used to describe $\phi_a$ in terms of $\phi_2(\equiv N$) and
\bref{2deq3} becomes the equation of motion for N,
\bea
\phi_a&=&-le_a^\mu\partial_\mu N,\\
0&=& (\nabla_\mu\nabla_\nu-g_{\mu\nu}\Box)N+g_{\mu\nu}N/l^2.
\label{2deq6}
\eea
\bref{2deq5} and \bref{2deq6} are exactly the same ones derived from \bref{Jackiw}.
The canonical form of \bref{2dG2} was also given by \cite{Fukuyama2} and two generators satisfy the conformal algebra without central charge
\bea
\{{\cal H}_\bot^f,{\cal H}_\bot^g\}&=&\{{\cal H}_1^f,{\cal H}_1^g\}={\cal H}_1^h,\nonumber\\
\{{\cal H}_\bot^f,{\cal H}_1^g\}&=&{\cal H}_\bot^h,
\eea
where
\bea
h&=&f\partial_1g-g\partial_1f,\nonumber\\
{\cal H}_\bot^f&=&\int dx^1f{\cal H}_\bot,~etc.
\eea
The explicit forms of ${\cal H}_\bot,~{\cal H}_1$ are given in \cite{Fukuyama2} and represent two dimensional diffeomorphism generators after gauge fixing
\be
e_{\mu a}=e^\chi\delta_{\mu a}.
\ee

The gauge part of this SO(2,1) is
\bea
{\cal P}_2(gauge)&=&-\frac{1}{2\pi}\mbox{Tr}\epsilon^{\mu\nu}F_{\mu\nu}\nonumber\\
&=&\partial_\mu{\cal C}_2^\mu
\label{1D1}
\eea
with
${\cal C}_2^\mu=-\frac{1}{2\pi}\mbox{Tr}\epsilon^{\mu\nu}A_\nu$ and
\be
{\cal C}_2^{(0)}(A)=\frac{1}{2\pi}\mbox{Tr}A.
\label{1D2}
\ee
%In one dimensional spacetime (relativistic particle), there is no vector.
%So in place of \bref{1D1} and \bref{1D2}
%2-cochain ${\cal C}_4^{(2)}$ defined in (2.23) plays the role.

%\section{Implications of local SO(2,d-1)}
%We have introduced the tetrad as \bref{tetrad1}.
%$Z_A$ a, however, may depend on d dimensional spacetime, and after the gauge fi%xing of \bref{GF1}
%\be
%Z_5(x)=l+\tilde{Z}_5(x),~~Z_a(x)=\tilde{Z}_a
%\ee
%and 
%\bea
%D_\mu Z_a&=&e_{\mu a}+D_\mu\tilde{Z}_a\\
%D_\mu Z_5&=&D_\mu\tilde{Z}_5
%\eea
%Corresponding to this situation, \bref{action1} has the additional terms
%\be
%I=\int d^4x \epsilon^{ABCDE}\epsilon^{\mu\nu\rho\sigma}\left[(\tilde{Z}_A/l)R_{%\mu\nu BC}R_{\rho\sigma DE}/(16g^2)+...\right]
%\ee
%Here the deviation from \bref{GF1} is small $\tilde{Z}_A\ll l$.

\section{Discussions}

We have argued that gravity in d dimensional spacetime is formulated as the gauge theory of SO(2,d-1).
The same but global symmetry is shared with conformal field theory in d-1 dimensional flat spacetime.
AdS/CFT correspondence has been discussed in the framework of nonsusy local field theory.
In both gravity and CFT, the Chern-Pontryagin index ${\cal P}_{2n}$, especially ${\cal P}_4$, play a crucial role.
In ${\cal P}_4(gravity)$, Einstein gravity, linear term in the Riemannian tensor, survives by virtue of scale violation. Whereas, the Lagrangian of gauge part appears as the dimensional descents $0,1-$cochains of ${\cal C}_4(gauge)$ + kinetic terms in three, two dimensions, respectively.

Gravitational Lagrangian has the surface term and higher derivative terms for $d\geq 4$, which may change the boundary condion of BH solution and affect Reisner-Nordstrom and Kerr/CFT correspondence \cite{Strominger}. More concretely speaking, Strominger et.al. identified near event-horizon extreme Kerr with $CFT_2$ by the central charge \cite{Strominger}. If the surface term appeared in our theory modifies the central charge, this correspondence may be affected. These are the soliton solution of gravity.  

Gauge theory of gravitation allows the other kind of soliton solution.
Let us consider the case of $d=5$ case. we adopted the gauge \bref{GF2}.
However we may set the kink solution
\bea
Z_A&=&(0,0,0,0,0,l)~~\mbox{at}~~x^5\subset (0,\infty) \nonumber\\
Z_A&=&(0,0,0,0,0,-l)~~\mbox{at}~~x^5\subset (-\infty,0) 
\eea
In this case the solution has the kink of step function at $x^5=0$.
This may be interpreted 3 dimensional brane.
Of course this is too simplified and we will discuss the detail of this process
in separate form.

Lastly we comment on the conformal invariance of the gravity.
For gravity, conformal transformation takes the form
\be
v_{\mu;\nu}+v_{\nu;\mu}=\frac{2}{d}g_{\mu\nu}
\label{CF2}
\ee
in place of \bref{CF1}. Here $v_{\mu;\nu}$ implies the covariant derivative.
The invariance under \bref{CF2} is recovered by the conformal (Weyl) tensor
\be
C^{\mu\nu\rho\sigma}=\mathring{R}^{\mu\nu\rho\sigma}-\frac{1}{d-2}\left(g^{\mu\rho}S^{\sigma\mu}-g^{\mu\sigma}S^{\rho\nu}-g^{\nu\rho}S^{\sigma\nu}+g^{\nu\sigma}S^{\rho\mu}\right)
\ee
with
\be
S^{\mu\nu}=\mathring{R}^{\mu\nu}-\frac{1}{2(d-1)}g^{\mu\nu}\mathring{R}.
\ee
This tensor vanishes iff the spacetime is conformally flat.
In $d=3$, this tensor vanishes identically but not all three dimensional spacetime is conformally flat. So this tensor does not characterize conformal flatness in $d=3$.  The "Weyl" tensor in $d=3$ is given by ${\cal C}_4^{(0)}(A)$ of \bref{CS1}
with replacement of $A$ by the Christoffel symbol,
\be
S_{CS}(gravity)=\int d^3x{\cal C}_4^{(0)}(\Gamma)=-\frac{1}{8\pi^2}\int d^3xTr\epsilon^{\nu\rho\sigma}\left(\Gamma_{\nu\beta}^\alpha\partial_\rho\Gamma_{\nu\alpha}^\beta+\frac{2}{3}\Gamma_{\nu\beta}^\alpha\Gamma_{\rho\lambda}\beta\Gamma_{\sigma\nu}^\lambda\right)
\ee
The new Weyl tensor $C^{\mu\nu}$ is given by the variation \cite{Jackiw3}
\be
\delta S_{CS}=\frac{1}{4\pi^2}\int d^3x \sqrt{g}C^{\mu\nu}\delta g_{\mu\nu}
\ee
and
\be
C^{\mu\nu}=\frac{1}{2\sqrt{g}}\left(\epsilon^{\mu\rho\sigma}R^\nu _{\sigma;\rho}+\epsilon^{\nu\rho\sigma}R^\mu _{\sigma;\rho}\right).
\ee
Thus the conformally flat transformation in gauge theory and conformal transformation in gravitation had some formal correspondence in the same dimension.
However, in the real world, the gravitation breaks the scale invariance and new kind of correspondence, $AdS_d/CFT_{d-1}$, appears.
%Of course this ${\cal C}$ is topological object and we must add dynamical field%, which is out of scope this paper \cite{Maldacena2}.

\section*{Acknowledgments}
We would like to thank Y. Sugawara, N. Ikeda, K. Kamimujra, N. Ogawa and A.Randono for very useful conversations. 
This work is supported in part by the grant-in-Aid 
for Scientific Research from the Ministry of Education, 
Science and Culture of Japan (No. 20540282).

\end{document}